\documentclass[twocolumn, prl, amssymb, superscriptaddress, aps, showpacs,preprintnumbers,
amsmath,showkeys,floatfix]{revtex4-1}

\usepackage{epstopdf}
\usepackage{graphics}
\usepackage{graphicx}
\usepackage{dcolumn}
\usepackage{bm}
\usepackage{longtable}
\usepackage{epsfig}
\usepackage{times}
\usepackage{url}
\usepackage{color}
\usepackage{subfigure}

\begin{document}

\title{Storage and manipulation of single x-ray photons via nuclear hyperfine splitting}

\author{Guangru \surname{Bai}}
\affiliation{Department of Physics, National University of Defense Technology, 410073 Changsha, China}

\author{Zengxiu \surname{Zhao}}
\affiliation{Department of Physics, National University of Defense Technology, 410073 Changsha, China}

\author{Jianpeng \surname{Liu}}
\affiliation{Department of Physics, National University of Defense Technology, 410073 Changsha, China}

\author{Zuoye \surname{Liu}}
\affiliation{School of Nuclear Science and Technology, Lanzhou University, 730000 Lanzhou, China}

\author{Guangyue \surname{Hu}}
\affiliation{CAS Key Laboratory of Geospace Environment and Department of Engineering and Applied Physics, University of Science
and Technology of China, Hefei, Anhui 230026, China}
\affiliation{CAS Center for Excellence in Ultra-intense Laser Science (CEULS), Shanghai 200031, China}


\author{Xiangjin \surname{Kong}}
\email{kongxiangjin@nudt.edu.cn}
\affiliation{Department of Physics, National University of Defense Technology, 410073 Changsha, China}

\date{\today}

\begin{abstract}

\end{abstract}

\begin{abstract}

We introduce a technique to store and manipulate single x-ray photons which relies on dynamically controlled absorption via nuclear hyperfine magnetic splitting. This scheme is inherently suitable for storage, on-demand generation and dynamical manipulation of single x-ray photons, for instance, the manipulation of the temporal shape, temporal splitting, the interference between x-ray photons and the control of the polarization. Our approach opens up new paths in x-ray quantum information.

\end{abstract}

\maketitle

The commissioning of novel x-ray sources opens the new field of x-ray quantum optics, which expands the light-matter interactions into x-ray-nuclear regime \cite{adams2013}. Compared to optical photons, x-rays have a number of desirable properties such as deeper penetration, better focus, and correspondingly superior spatial resolution, as well as robustness. Based on x-ray-nuclear interfaces, many coherent control tools of x-ray photons have been demonstrated experimentally \cite{rohlsberger2010collective,rohlsberger2012electromagnetically,PhysRevLett.111.073601,vagizov2014coherent,heeg2015tunable,heeg2015interferometric,haber2016,haber2017rabi,heeg2017spectral}, which provide potential applications for the fields of metrology, material science, biology and chemistry. Moreover, the properties of x-ray photons provide some advantages over the optical photons in quantum information technologies, for instance, x-ray photons are no longer limited by an inconvenient diffraction limit as for low-frequency photons and x-rays are resonant to nuclear transitions with long coherence times, which are very well isolated from the environment even at room temperature.

Storage and retrieval of flying photons on demand are key elements for quantum information technologies, which has been realized in optical regime \cite{almeida2004all,julsgaard2004experimental,lvovsky2009optical,specht2011single,steger2012quantum,maxwell2013storage,reiserer2015cavity}. Some proposals to store single x-ray photons have been put forward, for instance, the storage of narrow band x-ray pulses based on electromagnetically induced transparency (EIT) \cite{kong2016stopping}, the generation of x-ray photon echos using gradient frequency comb (GFC) and stepwise gradient echo (SGE) \cite{zhang2019nuclear,yeh2019spectral}. Similar photon-echo quantum memory techniques based on controlled reversible inhomogeneous broadening (CRIB) have been widely investigated in optical regime \cite{moiseev2004photon,kraus2006quantum,tittel2010photon}. Coherent storage of nuclear excitation (not the input x-ray pulse) has also been proposed and realized experimentally \cite{liao2012coherent,PhysRevLett.77.3232, smirnov1996nuclear}. Very recently, an optical memory protocol based on  Autler-Townes splitting (ATS) has been proposed \cite{PhysRevLett.113.123602} and realized experimentally \cite{saglamyurek2018coherent,PhysRevResearch.1.022004,rastogi2019discerning} which can not be transferred directly to the hard x-ray regime due to the lack of two-color x-ray lasers. In this manuscript, we propose a mechanism to store and manipulate a broadband single x-ray photon pulse via dynamically controlled nuclear hyperfine magnetic splitting (NHMS). Our protocol relies on controlled absorption through NHMS peaks and our results show highly effective storage and on-demand generation of broadband single x-ray photon pulses. Unlike the usual photon-echo quantum memory technique \cite{moiseev2004photon,kraus2006quantum,tittel2010photon}, our protocol uses a dynamically controlled NHMS to replace the requirement of CRIB. Compared with established theoretical proposals \cite{zhang2019nuclear,yeh2019spectral} for a broadband single x-ray photon pulse, our mechanism does not require the inherent high-speed operation of the mechanical motion in the original storage schemes which offers relaxed requirements for future experimental implementation. Moreover, coherent control of the photon echo based on CRIB have been proposed \cite{moiseev2010temporal,moiseev2013scalable} and realized \cite{sparkes2012precision}, which is very important in quantum information. Here, we present the methods to manipulate the signal x-ray photons, for instance, temporal shaping and splitting of retrieval x-ray pulses, interfering x-ray photons and polarization switching. This provides suggest applications in developing new quantum information technologies in x-ray regime such as x-ray qubits. 

\begin{figure}
	\includegraphics[width=0.5\textwidth]{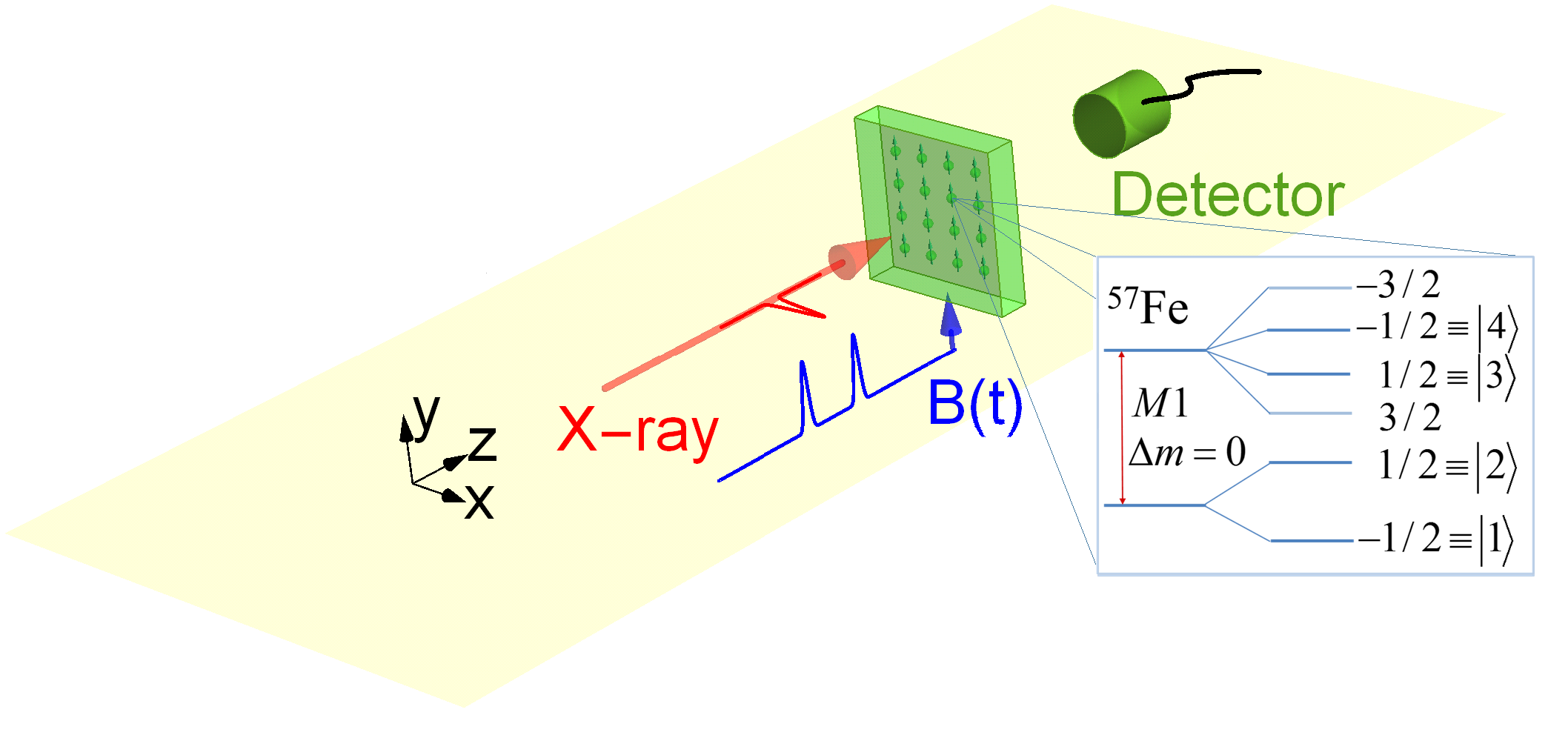}
	\caption{\label{system} Schematic of the nuclear forward scattering system with a nuclear ensemble containing $^{57}\mathrm{Fe}$. An incident x-ray pulse with linear polarization penetrates the resonant nuclear sample under a time-dependent magnetic field. The diagram of the nuclear levels is shown.}
\end{figure} 
We begin with a typical nuclear forward scattering system with monochromatized x-ray pulses shining along $z$ axis which is perpendicular to a sample containing M\"ossbauer nuclei $^{57}\mathrm{Fe}$, see Fig.~\ref{system}. This isotope has a stable ground state and a first excited state at 14.413 keV, corresponding to a wavelength of 0.86 $\AA$. These two states are connected via a magnetic dipole ($M1$) transition. With the nuclear resonance of approximate 4.66 neV natural linewidth, even when tuned to the nuclear transition energy, synchrotron pulses will act as broadband sources, with just one resonant photon in each pulse at most. In the presence of hyperfine magnetic field, the stable ground state of $^{57}\mathrm{Fe}$ (nuclear spin $I_{g}=1/2$) is then split into a doublet with $m_{g}=\pm 1/2$ and the first excited state (nuclear spin $I_{e}=3/2$, lifetime $\tau_0=141\text{ns}$) into a quadruplet with $m_e=\pm 1/2,\,\pm 3/2$. The hyperfine levels are coupled by six transitions, depending on the magnetic field geometry and polarization of the incident x-ray pulses. In the following, we first consider the x-ray pulse is linearly polarized with $\text{x}$-polarized light denoted as $\pi$-polarization by convention. A time-dependent magnetic field {\bf B}(t) that sets the quantization axis for the nuclear ground and excited state spin projections  $m_g$ and $m_e$ is parallel to the $\text{y}$ axis. 
In this scenario, the two $\Delta m=m_e-m_g=0$ magnetic dipole transitions will be driven by the incident pulse. 

The simplified Maxwell-Bloch equations in perturbation regime where the amplitude of probe Rabi frequency $\Omega_{0}\ll\Gamma$ describe the dynamics of x-ray-nuclear interaction system, which has been verified in nuclear forward scattering \cite{liao2012coherent,Kong_2014,PhysRevLett.113.123602}
\begin{align}
{{\partial }_{t}}{{\rho }_{S}}(z,t)=&-\frac{\Gamma}{2}{{\rho }_{S}}(z,t)-i\Delta (t){{\rho }_{P}}(z,t) \, , \label{equ1} \\ 
{{\partial }_{t}}{{\rho }_{P}}(z,t)=&-\frac{\Gamma}{2}{{\rho }_{P}}(z,t)-i\Delta (t){{\rho }_{S}}(z,t)+i\frac{C}{2}{{\Omega }_{p}(z,t)}\, ,\label{equ2}\\
\frac{1}{c}\partial_t\Omega_{p}(z,t)&+{{\partial }_{z}}{{\Omega }_{p}(z,t)}=i\frac{\beta}{C}\rho_{P}(z,t) \, , 
\label{equ3}
\end{align}
where 
\begin{align}
&\Delta (t)=\frac{1}{\hbar }\left({m}_{e4}{{\mu }_{e}}-{{m}_{g1}}{{\mu }_{g}}\right){{\mu }_{{N}}}{B}(t)\, , \\
&{{\rho }_{P}}(z,t)={{\rho }_{41}}(z,t){+}{{\rho }_{32}}(z,t)=2i\text{Im}[{\rho }_{41}(z,t)]\, , \\
&{{\rho }_{S}}(z,t)={{\rho }_{41}}(z,t){-}{{\rho }_{32}}(z,t)=2\text{Re}[{\rho }_{41}(z,t)]\, .
\end{align}

In the above equations, the states $|1\rangle$ and $|2\rangle$ denote the  two ground states with $m_g=-1/2$ and $m_g=1/2$, respectively, and $|3\rangle$ and $|4\rangle$ the two excited states with $m_e=1/2$ and $m_e=-1/2$, respectively. ${{\mu }_{{N}}}$ is nuclear magneton, ${\mu }_{g}$ and ${\mu}_{e}$ are the Land\'e factors of the ground states and excited states, respectively. The constant $\Gamma$ is the spontaneous decay rate of $^{57}\mathrm{Fe}$, $C=\sqrt{2/3}$ is the corresponding Clebsch-Gordan coefficient for the two $\Delta m=0$ transitions in Fig.~\ref{system}, and $\beta=4\Gamma\xi/L$ where $\xi$ is the  resonant thickness and $L$ the thickness of nuclear target. $\rho_{P}(z,t)$ and $\rho_{S}(z,t)$ denote nuclear polarization and spin coherences, respectively.

With the goal of highly reliable storage and retrieval of a broadband single x-ray photon pulse, we investigate the configuration that pulsed magnetic fields are applied.  As an example, we consider a Gaussian input single x-ray photon pulse with 9 ns bandwidth and 14.413 keV central frequency which is resonant to the nuclear transition of $^{57}\mathrm{Fe}$. In our protocol, we first consider the magnetic field has the same temporal shape as the input x-ray photon with the condition $\text{A}_{B}=\int\Delta(t)dt\approx\pi$ which has been adopted in recent works \cite{moiseev2017multiresonator,bao2021demand}. The first magnetic field pulse arrives simultaneously as the input x-ray pulse. Here, $\text{A}_{B}$ is defined as the area of the magnetic field which is set to $\pi$ for all the numerical simulations presented in this manuscript. Under the above conditions, an analytical solution of Eq.~\ref{equ1}-~\ref{equ3} with the first pulsed magnetic field is derived with some approximations (details are presented in Supplemental Materials) 

\begin{align}
&{{\rho }_{S}}(L,t)=A_{01}\left(1-\text{Cos} \left[\int_{{{t}_{01}}}^{t}{{{\Delta }_{1}}(t')dt'}\right]\right){{e}^{-\frac{\Gamma }{2}t}}\, , \label{f1} \\ 
&{{\rho }_{P}}(L,t)=A_{01}\text{Sin} \left[\int_{{{t}_{01}}}^{t}{{{\Delta }_{1}}(t')dt'}\right]{{e}^{-\frac{\Gamma }{2}t}}\, , \label{f2} \\ 
&{{\Omega }^1_{p}}(L,t){=}{{\Omega }_{p}}(0,t)-\Omega_L\text{Sin}\left[\int_{{{t}_{01}}}^{t}{{{\Delta }_{1}}(t')dt'}\right]{{e}^{-\frac{\Gamma }{2}t}}\, , \label{f3}
\end{align}
where $A_{01}=\frac{{C{\Omega }_{0}}}{4{{\Delta }_{01}}}{{e}^{-\frac{2\Gamma\xi}{\Delta_{01}}}}$ and $\Omega_L={{\Omega }_{0}}(1-{{e}^{-\frac{2\Gamma\xi}{\Delta_{01}}}})$. ${\Omega }^1_{p}$ is Rabi frequency of the output x-ray photon, $\Delta_{1}(t)$ and $\Delta_{01}$ is the temporal splitting and the splitting amplitude caused by the first magnetic field, and $t_{01}$ is the starting time.
\begin{figure}[t]
	\centering
	\subfigure[\mbox{ } Storage and retrieval of a broadband single x-ray photon pulse. ]{
		\label{r1a} 
		\includegraphics[width=3.1in]{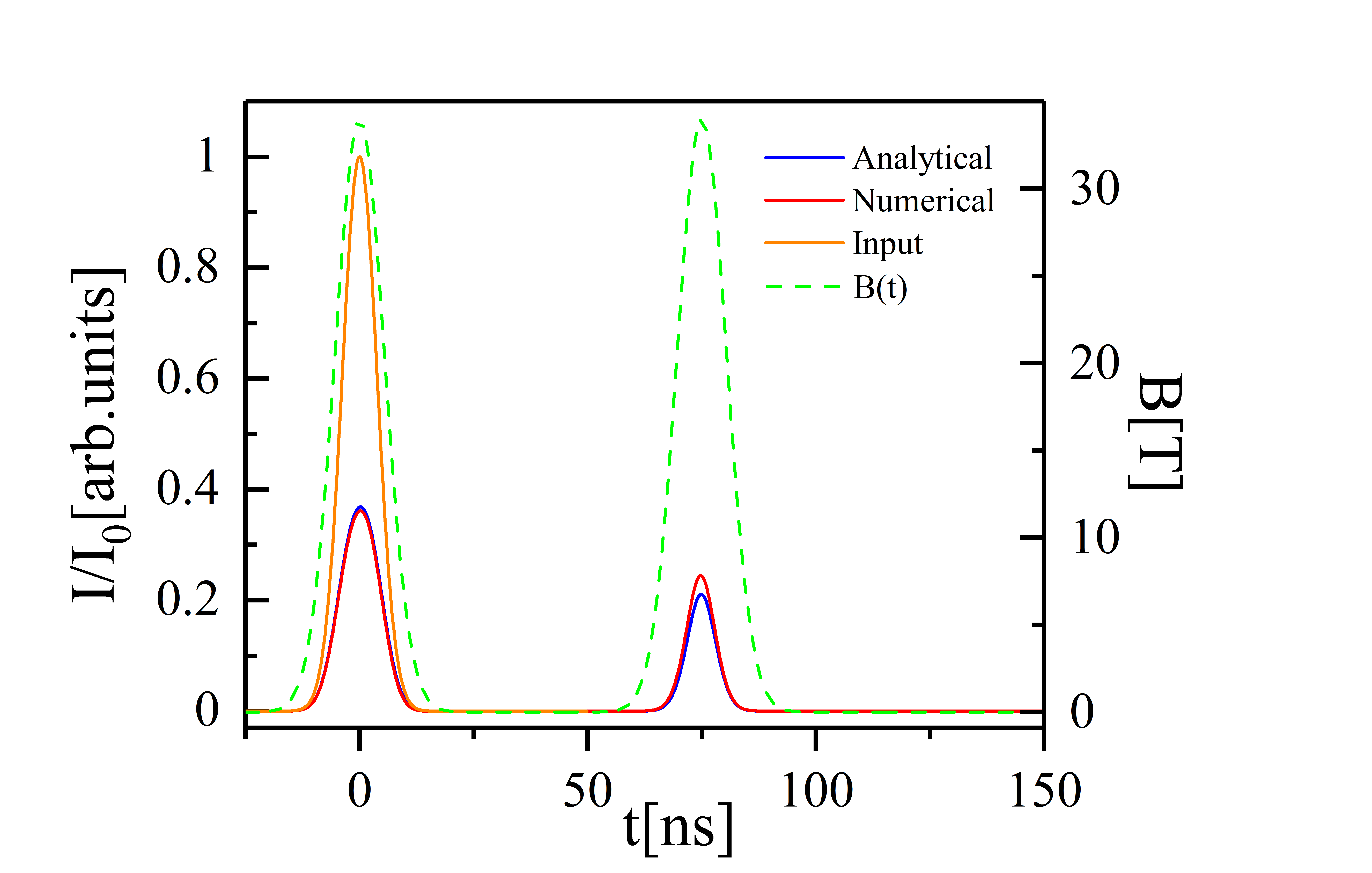}}
	\hspace{1in}
	\subfigure[\mbox{ } Efficiency of the first echo depending on the resonant thickness. ]{
		\label{r1b} 
		\includegraphics[width=3.1in]{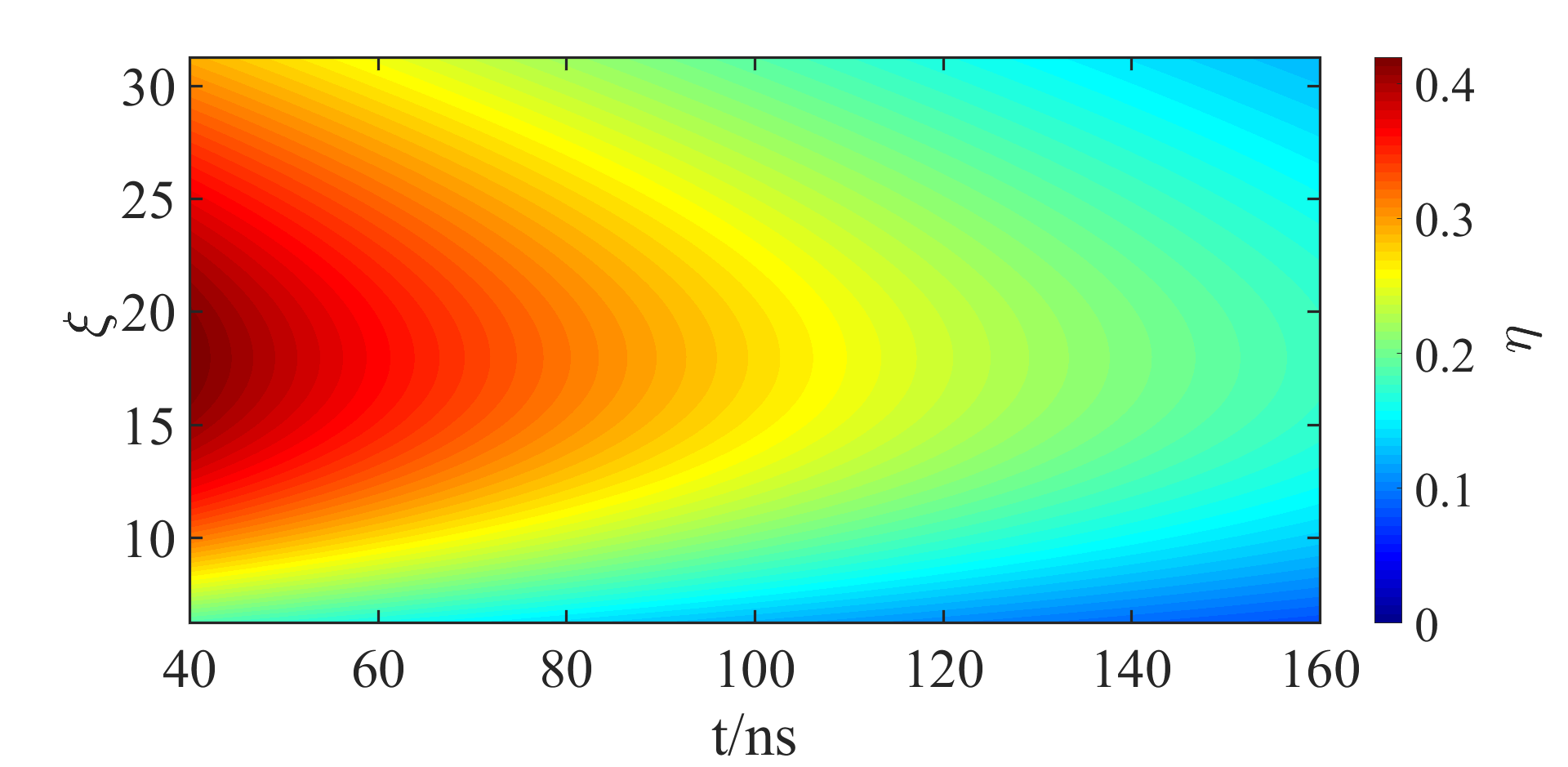}}
	\caption{(a) Storage and retrieval of a broadband single x-ray photon pulse. The input x-ray pulse (orange line) shines the nuclear sample together with a pulsed magnetic field (dashed green line). The solid red and blue lines are the numerical and analytical results for the propagated x-ray photon, respectively. The x-ray photon is recovered by the second pulsed magnetic field at the central time $t=75$ ns. (b) The efficiency of the first echo depending on the resonant thickness. Here, the spontaneous decay is taken into account and makes the efficiency decrease with time.
	}	
\end{figure}

When the single x-ray photon pulse arrives with the magnetic field, it is absorbed by the nuclei and transferred to the nuclear polarization coherence $\rho_{P}$ and nuclear spin coherence $\rho_{S}$ inside the medium. After some time, the output x-ray photon leaves out the nuclear medium meanwhile the polarization coherence $\rho_{P}$ evolves to zero since $\int\Delta(t)dt\approx\pi$, see Eq.~\ref{f2} and Eq.~\ref{f3}. Surprisingly, the nuclear spin coherence term $\rho_{S}$ reaches a maximum which is preserved if we neglect the spontaneous decay $\Gamma$, see Eq.~\ref{f1}. The numerical results are shown in Fig.\ref{r1a}. After a chosen storage time $T_1=75$ ns at random, the second magnetic field is switched on as a read-out signal. The analytical solution under some approximations is presented in the following
\begin{align}
{{\rho }_{S}}(L,t)&=2A_{02}\left(1-\frac{2\Gamma\xi}{\Delta_0}\right)\text{Cos} \left[\int_{{{t}_{02}}}^{t}{{{\Delta }_{2}}(t')dt'}\right]{{e}^{-\frac{\Gamma }{2}t}} \nonumber \\
&\,\,\,\,\,\,+2A_{02}\frac{2\Gamma\xi}{\Delta_0}{{e}^{-\frac{\Gamma }{2}t}}\, , \label{s1} \\ 
{{\rho }_{P}}(L,t)&=-2A_{02}\left(1-\frac{2\Gamma\xi}{\Delta_0}\right)\text{Sin}\left[\int_{{{t}_{02}}}^{t}{{{\Delta }_{2}}(t')dt'}\right]{{e}^{-\frac{\Gamma }{2}t}}\, , \label{s2} \\ 
{{\Omega }^2_{p}}(L,t)&=2{{\Omega }_{0}}\frac{2\Gamma\xi}{\Delta_{02}}{{e}^{-\frac{2\Gamma\xi}{\Delta_{02}}}}\text{Sin}\left[\int_{{{t}_{02}}}^{t}{{{\Delta }_{2}}(t')dt'}\right]{{e}^{-\frac{\Gamma }{2}t}}\, , \label{s3} 
\end{align}
where $A_{02}=\frac{{C{\Omega }_{0}}}{4{{\Delta }_{02}}}{{e}^{-\frac{2\Gamma\xi}{\Delta_{02}}}}$ and ${{\Omega }^2_{p}}$ represents the Rabi frequency of the first echo pulse. The shape of the regenerated x-ray pulse is approximately the same as input since $\text{Sin}\left[\int_{{{t}_{02}}}^{t}{{{\Delta }_{2}}(t')dt'}\right]\approx{\Delta }_{2}(t)/{\Delta }_{02}$ under the initial conditions. Based on the above equations, we find that the magnetic field mediates the transformation from the nuclear spin coherence $\rho_{S}$ to the nuclear polarization coherence $\rho_{P}$ and the x-ray photon. First, the x-ray photon is mapped to the nuclear spin and polarization coherences via the absorption in the presence of the magnetic field. Then the nuclear spin coherence $\rho_{S}$ is preserved which can be transformed back to the output x-ray photon by the read-out magnetic field. The retrieval of the x-ray photon is shown in Fig.~\ref{r1a} and the approximate analytical solution fits well with the numerical result. Our analytical method can also been applied in the optical domain. 
The storage mechanism discussed here is via the dynamically controlled absorption which is different from the narrow-band storage protocol based on EIT \cite{kong2016stopping} and this makes our approach suitable for storing a broadband single x-ray photon pulse. A detailed discussion of these two mechanisms is demonstrated in Ref.~\cite{saglamyurek2018coherent,rastogi2019discerning}.



The Rabi frequency of the regenerated x-ray photon using a $\pi$ read-out magnetic field pulse is dependent on the resonant thickness $\xi$, see Eq.~\ref{s3}. When $\frac{2\Gamma\xi}{\Delta_{02}}\sim 1$, the amplitude of Rabi frequency reaches a maximum which obtains a optimal echo efficiency. This efficiency is defined as 
\begin{equation}
\eta=\frac{\int\vert\Omega^2_p\left(L,t\right)\vert^2dt}{\int\vert\Omega_p\left(0,t\right)\vert^2dt}
\end{equation}
and the numerical result of $\eta$ as a function of the storage time and $\xi$ is demonstrated in Fig.~\ref{r1b}. Using the read-out magnetic field pulse with area $\pi$, our mechanism provides a maximal nearly $55\%$ storage efficiency of the first retrieved x-ray photon with the optimal $\xi=16 \left(\frac{2\Gamma\xi}{\Delta_{02}}=1\right)$ if the spontaneous decay is neglected. When the spontaneous decay is considered, the storage efficiency decreases with time, as shown in Fig~\ref{r1b}. A larger storage efficiency can be provided using magnetic fields with different areas.

\begin{figure}[h]
	\includegraphics[width=3.1in]{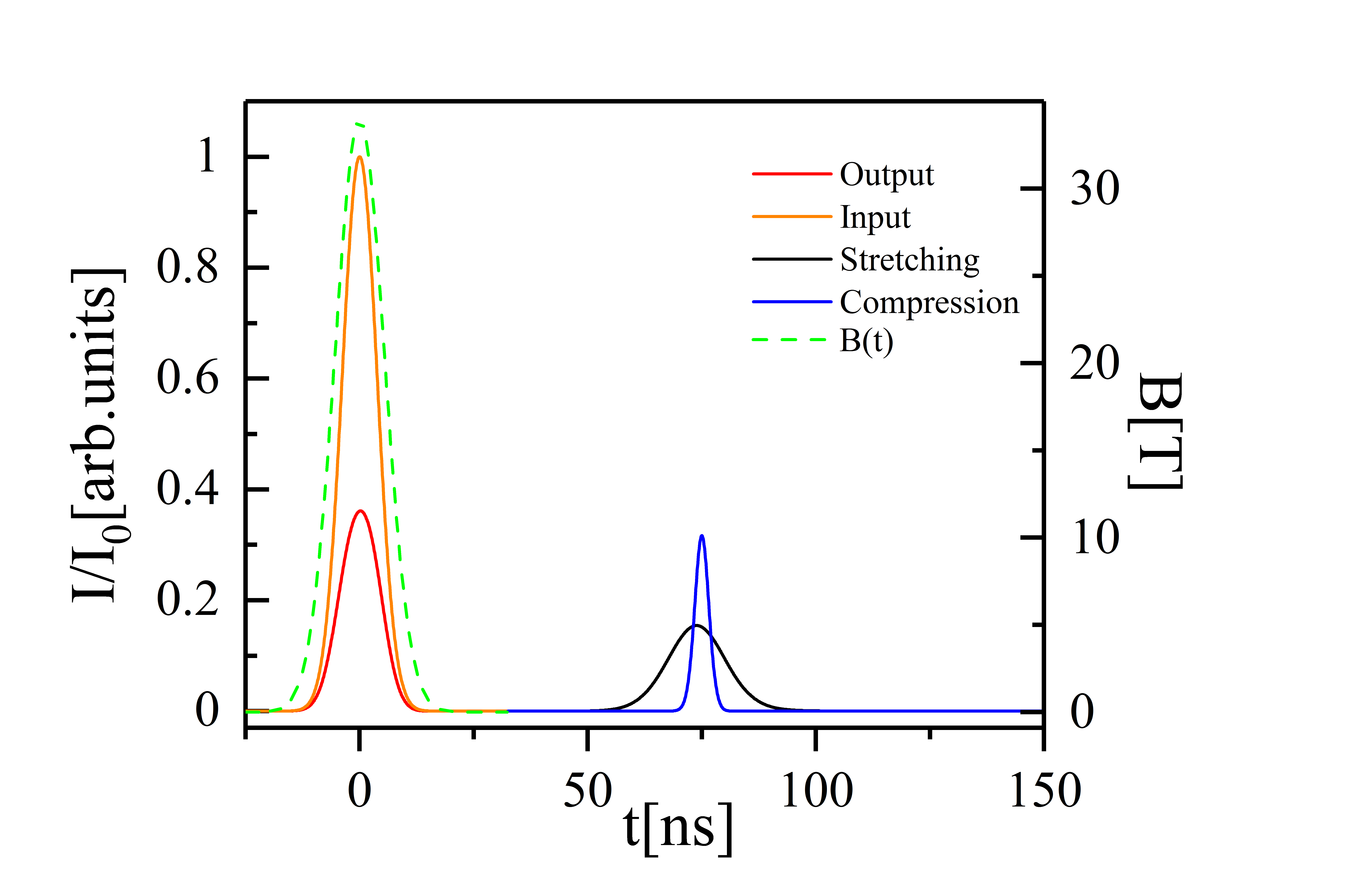}
	\caption{\label{r3} Temporal manipulation of regenerated x-ray photons. Two read-out magnetic field pulses with duration $4.5$ ns and $18$ ns are used, which are corresponding to the temporal compression(solid blue line) and stretching(solid black line), respectively.}
\end{figure} 
	

According to Eq.~\ref{s3}, the temporal shape and amplitude of the recovered x-ray photon can be manipulated by the read-out magnetic field. As in the writing process, we choose a magnetic field with area $\pi$ and matches the temporal shape of the input x-ray signal. Two pulsed magnetic fields with the same area $\pi$ but different durations which do not match the input x-ray signal are used in the read-out process for comparison. As presented in Fig.~\ref{r3}, the durations of the read-out magnetic fields are $4.5$ ns and $18$ ns, which are corresponding to the temporal compression and stretching, respectively. A larger amplitude of the recovered x-ray photon is obtained with the temporal compression. Our approach is robust to the bandwidth of the controllable incident few ns x-ray pulse \cite{vagizov2014coherent,liao2018} which determines the properties of the optimal magnetic field for storage. The magnetic field of dozens of Tesla with the duration time from a few nanoseconds to hundreds of nanoseconds has been widely used in laser-plasma experiments \cite{fujioka2013kilotesla,Santos_2015,sakata2018magnetized,bolanos2019highly,hu2020pulsed} and the nuclear forward scattering of synchrotron radiation in a pulsed magnetic field with a few milliseconds under the repetition rate of 6 min$^{-1}$ has been demonstrated \cite{strohm2010nuclear}, which make our protocol in experimental reach. 
\begin{figure}[t]
	\centering
	
	\subfigure[\mbox{ } Intensity of regenerated x-ray photons for $\xi=8$.]{
		\label{r4a} 
		\includegraphics[width=3.1in]{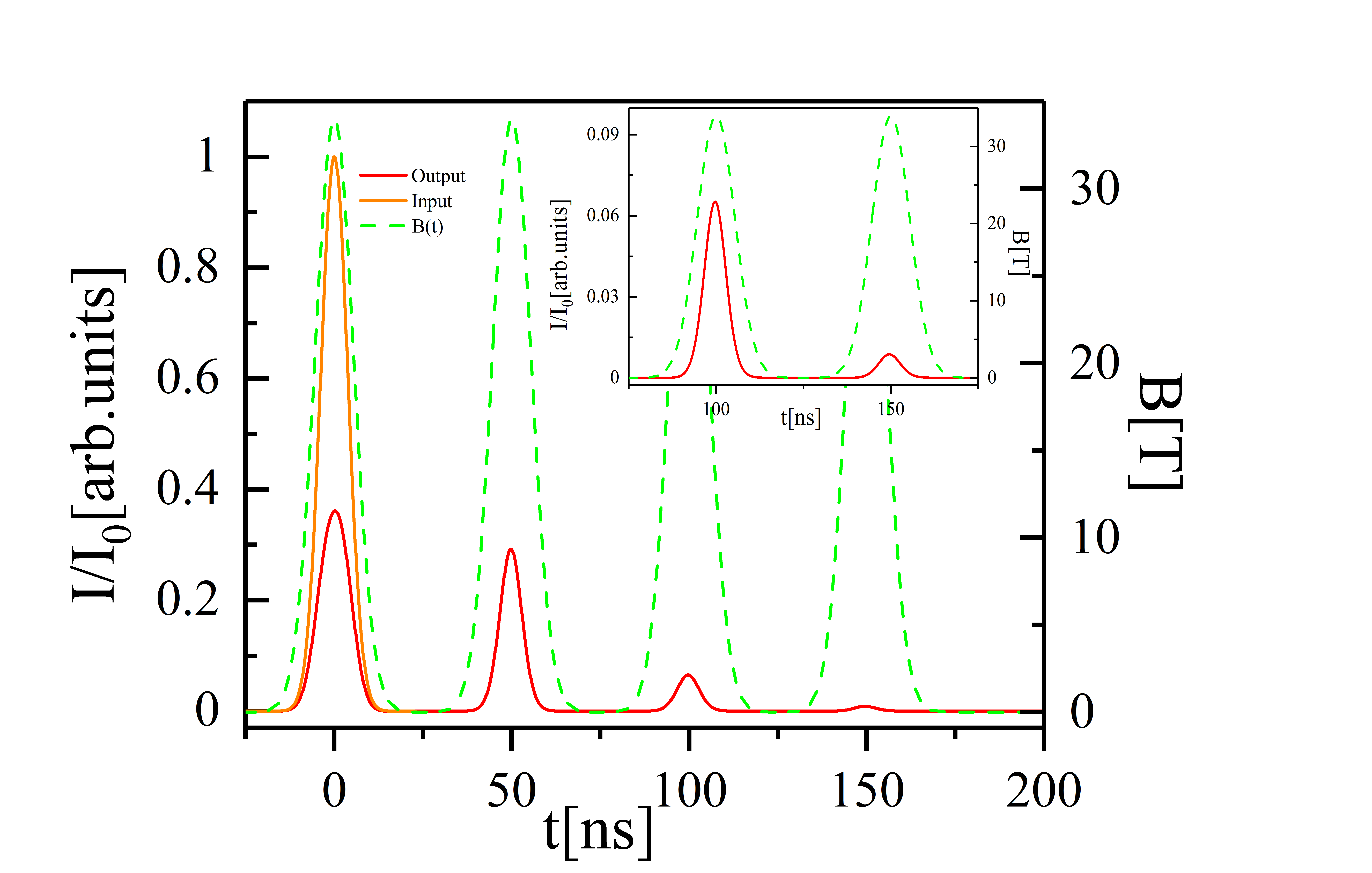}}
	\hspace{1in}
	\subfigure[\mbox{ } Rabi frequency of regenerated x-ray photons for $\xi=8$ and $\xi=24$.]{
		\label{r4b} 
		\includegraphics[width=3.1in]{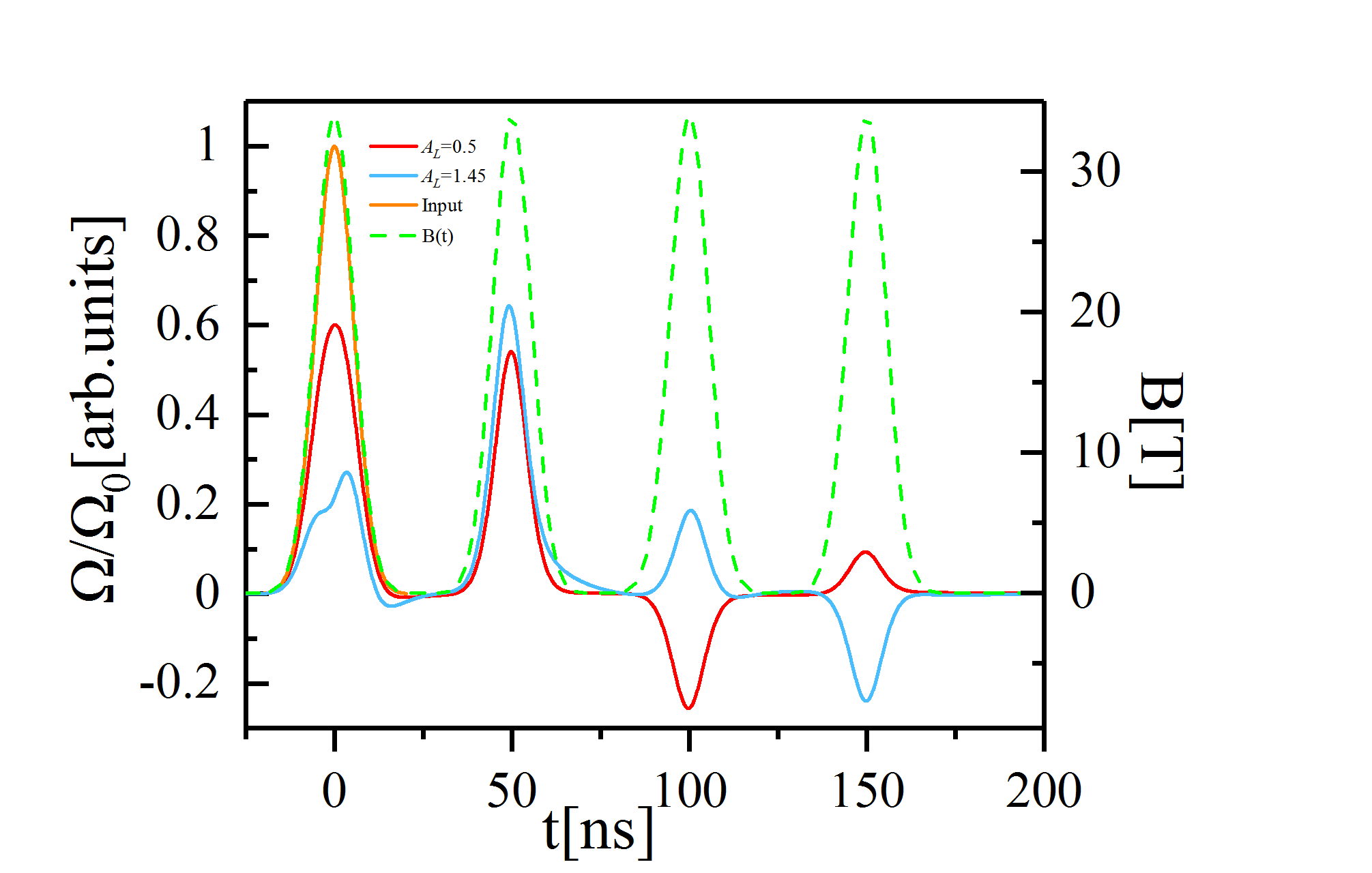}}
	\caption{Temporal splitting of regenerated x-ray signals. The four identical magnetic field pulses (dashed green lines) play the role of temporal beam-splitters which split the input x-ray pulses in time domain. 
	}
	\label{r4} 
\end{figure}

In our storage protocol, the signal is recalled by the magnetic field. Using a train of magnetic field pulses, multiple x-ray echoes can be  generated in time domain which makes the storage system as temporal x-ray beam-splitters. In our simulation, four identical magnetic field pulses with area $\pi$ are used and each of them splits the x-ray photon in time domain. The analytical solution for the Rabi frequency of the second x-ray photon echo is derived as
\begin{align}
 {\Omega^3_{p}}(L,t)=2{{\Omega }_{0}}\left(A_L^2-A_L\right){{e}^{-A_L}}\text{Sin}\left[\int_{{{t}_{03}}}^{t}{{{\Delta }_{3}}(t')dt'}\right]{{e}^{-\frac{\Gamma }{2}t}}\, ,
\end{align}
where $A_L=\frac{\beta L}{2\Delta_{03}}=\frac{2\Gamma\xi}{\Delta_{03}}$.
The numerical calculations for the intensities of the temporal split x-ray signals with $A_L=0.5(\xi=8)$ is demonstrated in Fig~\ref{r4a}. The temporal shape is approximately the same for each echo, but the sign of the amplitude of the Rabi frequency is different. Depending on $A_L$ is larger or smaller than one, $\Omega_p^3$ may have different signs. Fig~\ref{r4b} shows the temporal Rabi frequencies for $A_L=0.5$ and $A_L=1.45$ cases, in which there is a $\pi$ phase shift between the third and fourth echoes. And if $A_L\sim 1$, the intensity of the second echo will be very close to zero which is quite different from the case of the first echo.

\begin{figure}[]
	\centering
	
	\subfigure[\mbox{ } Intensity of regenerated x-ray photons for $\xi=8$.]{
		\label{r5a} 
		\includegraphics[width=3.1in]{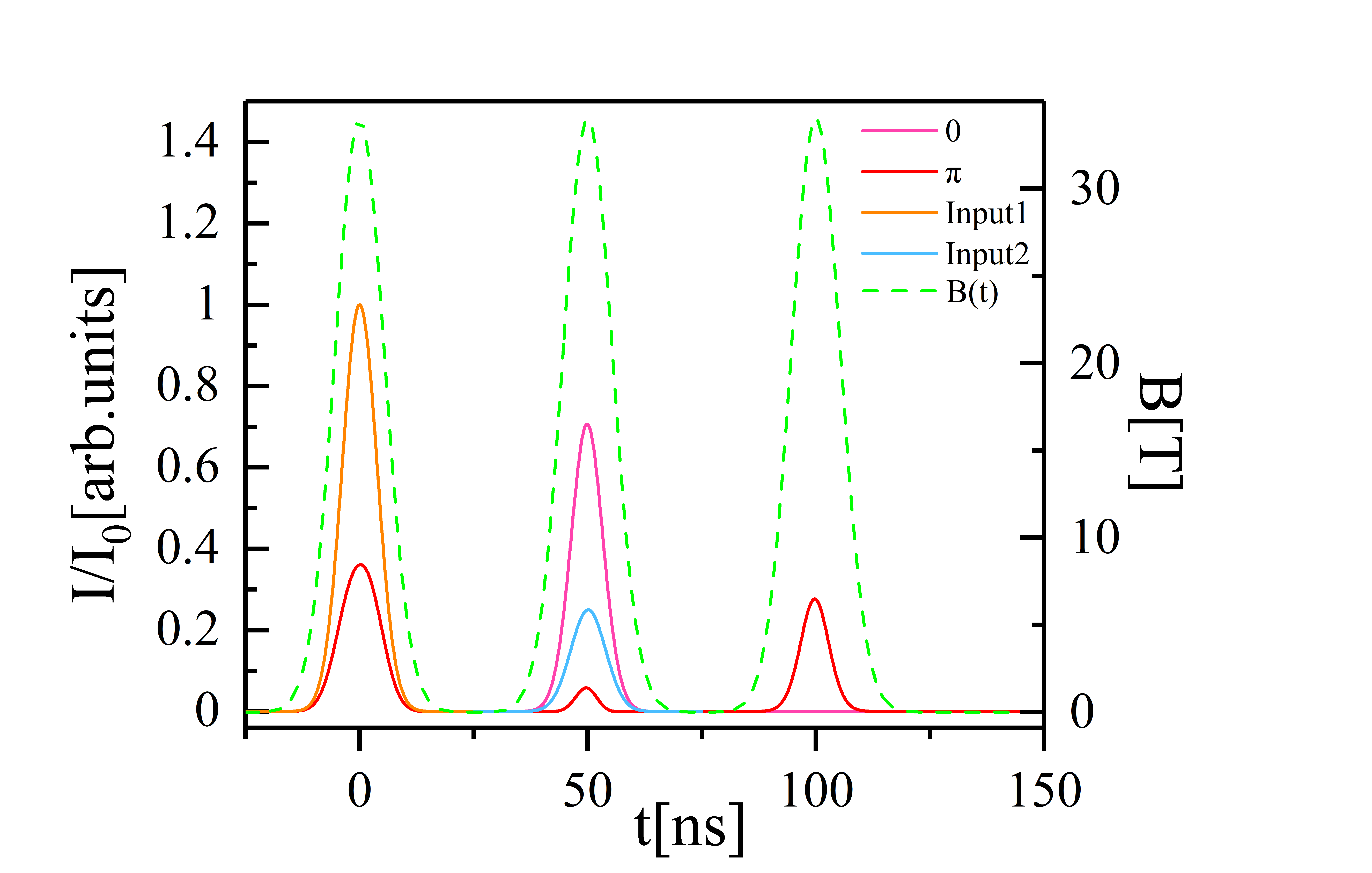}}
	\hspace{1in}
	\subfigure[\mbox{ } Intensity of regenerated x-ray photons for $\xi=24$.]{
		\label{r5b} 
		\includegraphics[width=3.1in]{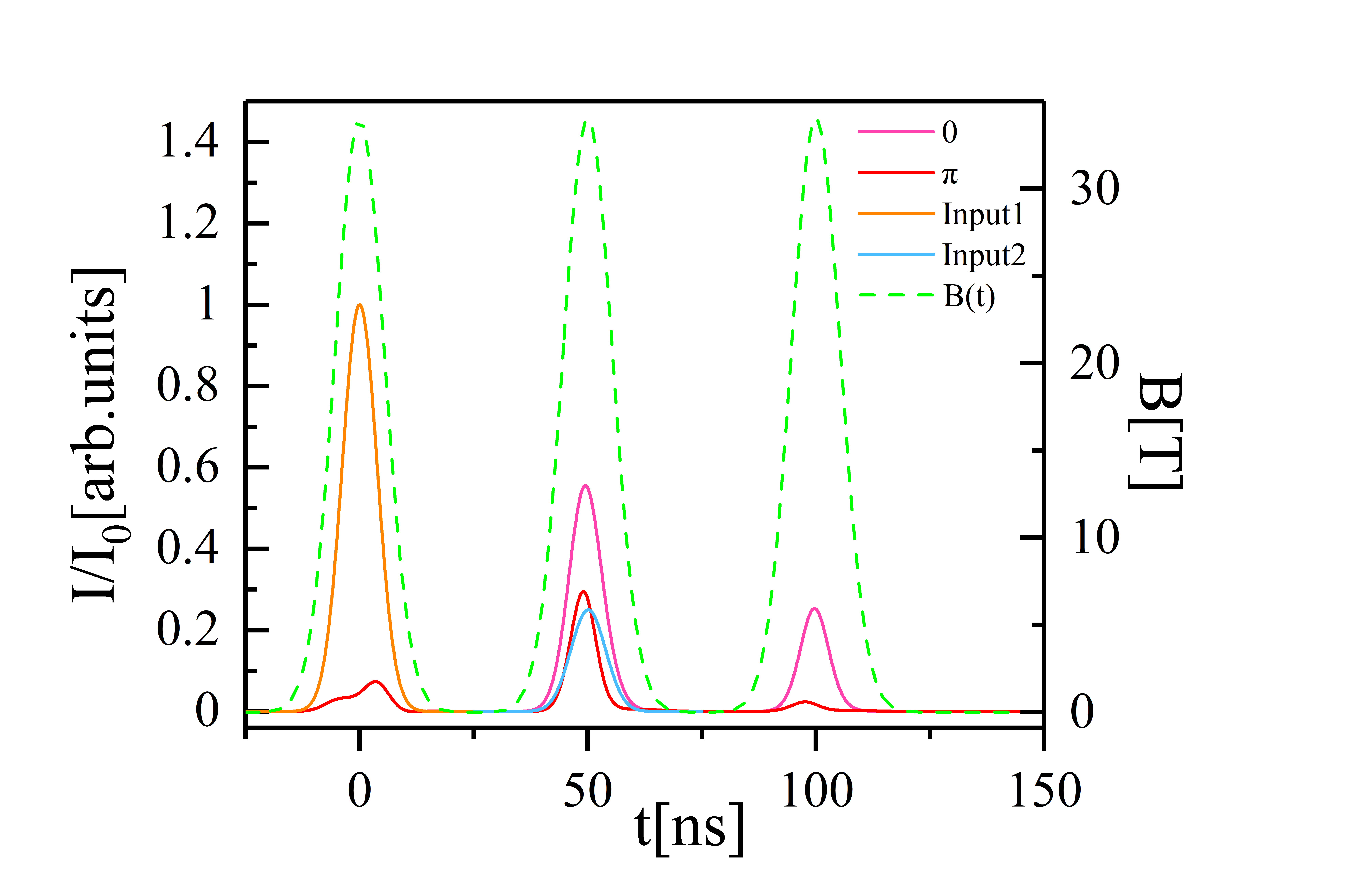}}
	\caption{Interference of two regenerated x-ray signals from different input modes. (a) and (b) show the numerical results for different resonant thickness $\xi$. Three identical magnetic field pulses are used(dashed green lines).  
	}
	\label{r5} 
\end{figure}

Depending on the phase shift of the echoes, our protocol can be used for operating the interference between x-ray photons recovered from distinct temporal inputs. The simulations for $A_L=0.5$ and $A_L=1.45$ have been performed, respectively. For each $A_L$, we consider two cases which are $0$ and $\pi$ phase shifts between the two input x-ray pulses. The numerical results are demonstrated in Fig.~\ref{r5}. Comparing with the first input x-ray pulse, the second one has the same temporal shape and a half amplitude of the Rabi frequency. The time internal is  $50$ ns and three identical $\pi$ magnetic field pulses are used. When $A_L=0.5$, the second echo shown in Fig.~\ref{r5a} is generated under the interference between the signals from the two input x-ray pulses. Destructive and constructive interferences occur for $0$ and $\pi$ phase shifts cases, respectively. Conversely, destructive and constructive interference occur when $A_L=1.45$ for $\pi$ and $0$ phase shifts cases, respectively, as presented in Fig.~\ref{r5b}. Our numerical results shows our protocol may provide a suitable platform for manipulating the interference between x-ray photons.

\begin{figure}[h]
	\includegraphics[width=3.0in]{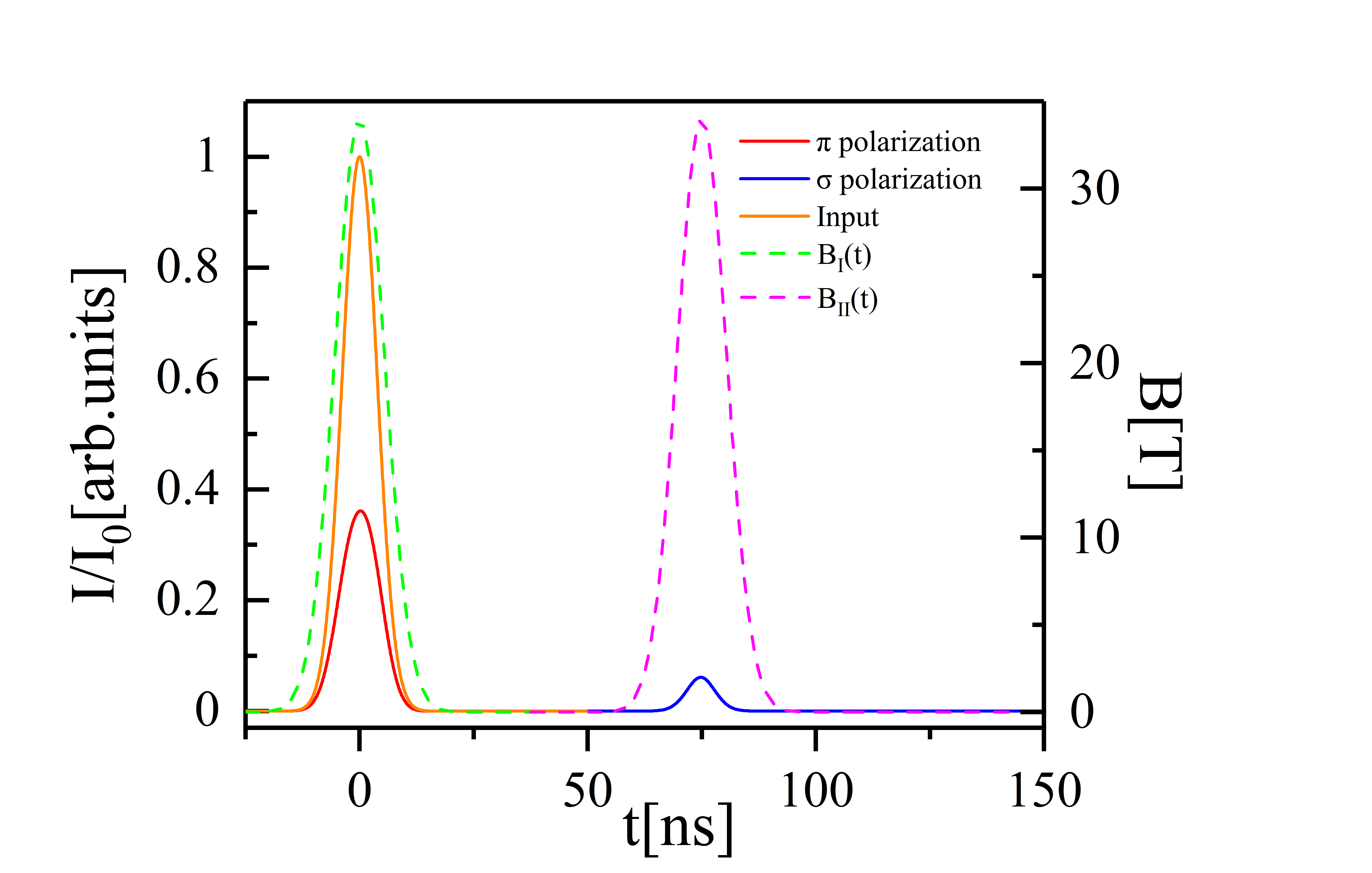}
	\caption{\label{r6} Polarization control of the x-ray echo. The initial x-ray pulse (orange solid line) is $\pi$-polarized and the magnetic field (green dashed line) is along $\text{y}$ direction. The output of x-ray pulse (red solid line) is $\pi$-polarized. After applying a subsequent magnetic field pulse parallel to $\text{x}$ direction (purple dashed line), a $\sigma$-polarized x-ray echo is generated (blue solid line). }
\end{figure}

In the discussions above, the polarization of x-ray echoes stays conserved since the directions of the magnetic fields keep constant. The situation changes if a subsequent magnetic field with a different direction is applied, which leads to the change of the quantization axis and a redistribution of the nuclear spin $\rho_{S}$ \cite{PhysRevLett.77.3232,palffy2009single}. Here we discuss the case that a $\pi$-polarization input x-ray pulse shines the nuclear sample with a pulsed magnetic field along $\text{y}$ direction as an example. A $\sigma$-polarization echo is generated with a subsequent pulsed magnetic field along $\text{x}$ direction, as shown in Fig.~\ref{r6}. The polarization of the x-ray echo is controlled by the direction of the subsequent magnetic field pulse. Our platform can be used to realize the single-photon entanglement and logical gates in keV regime proposed in Ref.~\cite{palffy2009single} and Ref.~\cite{gunst2016logical}, respectively. Compared with the original proposals\cite{palffy2009single,gunst2016logical}, our method works and provides a much higher efficiency due to the effective storage of x-ray signal, which makes a further step to achieve polarization-encoded x-ray qubits.

In conclusion, we have proposed a novel protocol for the storage and manipulation of a broadband single x-ray photon pulse based on dynamically controlled NHMS. The storage is deterministic and allows flexible storage times. Meanwhile, the temporal shape, phase and polarization of the recovered x-ray pulse can be controlled. Our approach points a way to realize x-ray photonics devices using nuclear transitions. We believe our work will bring new opportunities for x-ray quantum information and x-ray science.

\begin{acknowledgments}
XJK acknowledges the support by the National Natural Science Foundation of China (NSFC) under Grant No. 11904404. JPL acknowledges the support by the Innovative Research Program for Graduates of Hunan under Grant CX20200013. 
\end{acknowledgments}

\bibliography{storage}


\clearpage
\begin{widetext}
\begin{center}
{\LARGE \bf Supplementary Materials}
\end{center}
\renewcommand{\theequation}{S\arabic{equation}}
\renewcommand{\thesection}{S-\arabic{section}}
\renewcommand{\thefigure}{S\arabic{figure}}
\renewcommand{\thetable}{S\arabic{table}}
\renewcommand{\bibnumfmt}[1]{[S#1]}
\renewcommand{\citenumfont}[1]{S#1}
\setcounter{equation}{0}
\setcounter{figure}{0}

In the following we present the derivation of the analytical solutions under effective approximations which describe the dynamics of the storage and retrieval of single x-ray photons through a nuclear target containing $^{57}\mathrm{Fe}$.

\section{1. Maxwell-Bloch equations}
	
In the presence of the magnetic field,  the two $\Delta m=m_e-m_g=0$ transitions will be driven by the incident linearly polarized x-ray pulses. Maxwell-Bloch equations (MBEs) can be used to describe the dynamics of the system which are written as 
	\begin{align}
	\partial _t\rho _{11}=&\Gamma(C_{13}^2\rho _{33}+C_{14}^2\rho _{44})
	-\frac{i}{2}C_{14}(\Omega_{p}\rho _{14}-\Omega_{p}^*\rho_{41}) \, , \label{equ1} \\ 
	\partial _t\rho _{22}=&\Gamma(C_{23}^2\rho _{33}+C_{24}^2\rho _{44})
	-\frac{i}{2}C_{23}(\Omega_{p}\rho _{23}-\Omega_{p}^*\rho_{32}) \, , \label{equ2} \\ 
	\partial _t\rho _{32}=&-\frac{1}{2}(2i\Delta_{p,4\rightarrow2}+C_{13}^2\Gamma+C_{23}^2\Gamma)\rho _{32}
	-\frac{i}{2}C_{23}\Omega_{p}(\rho_{33}-\rho _{22}) \, , \label{equ3} \\ 
	\partial _t\rho _{33}=&-\Gamma(C_{13}^2+C_{23}^2)\rho _{33}
	+\frac{i}{2}C_{23}(\Omega_{p}\rho _{23}-\Omega_{p}^*\rho_{32}) \, , \label{equ4} \\ 
	\partial _t\rho _{41}=&-\frac{1}{2}(2i\Delta_{p,5\rightarrow1}+C_{14}^2\Gamma+C_{24}^2\Gamma)\rho _{41}
	-\frac{i}{2}C_{14}\Omega_{p}(\rho_{44}-\rho _{11}) \, , \label{equ5} \\ 
	\partial _t\rho _{44}=&-\Gamma(C_{14}^2+C_{24}^2)\rho _{44}
	+\frac{i}{2}C_{14}(\Omega_{p}\rho _{14}-\Omega_{p}^*\rho_{41})\, , \label{equ6}\\
	\frac{1}{c}\partial _t\Omega_{p}&+\partial _z\Omega_p=i{\beta}\left(\frac{\rho _{41}}{C_{14}}+\frac{\rho _{32}}{C_{23}}\right)\, . 
	\label{equ7} 
	\end{align}
	
	In the above equations, the states $|1\rangle$ and $|2\rangle$ denote the  two ground states with $m_g=-1/2$ and $m_g=1/2$, respectively, and the  two ground states $|3\rangle$ and $|4\rangle$ with $m_e=1/2$ and $m_e=-1/2$ respectively. The shortened notation used for the Clebsch-Gordan coefficients is $C_{ij}=C(I_g\,I_e\,  1;m_g\,  m_e\, M )$ where $i\in \{1,2\}$ sets the value of $m_g$ and $j\in \{3,4\}$ the one of $m_e$.
	Furthermore, $\Delta_{p,3\rightarrow2}=\omega_{32}-\omega$ and $\Delta_{p,4\rightarrow1}=\omega_{41}-\omega$, where $\omega_{41}$ and $\omega_{32}$ are the resonant frequencies of the $|1\rangle\rightarrow|4\rangle$ and $|2\rangle\rightarrow|3\rangle$ transitions,and $\omega$ is the resonant frequency of the ground state and the excited state in the absence of the magnetic field.$\Gamma$ is the spontaneous decay rate comprising the radiative and the internal conversion channel. $\beta=4\Gamma\xi/L$ where $\xi$ is the resonant thickness and $L$ the thickness of nuclear target.
	
	The initial conditions are considered as follows 
	\begin{align}
	\rho_{11}(z,0)=&0.5\, , \label{equ8} \\ 
	\rho_{22}(z,0)=&0.5\, , \label{equ9} \\ 
	\Omega_{p}(z,0)=&0\, , \label{equ10} \\ 
	\Omega_{p}(0,t)=&\Omega_0{{e}^{-2\ln 2{{(\frac{t-{{t}_{0}}}{\tau })}^{2}}}}\, . 
	\label{equ11} 
	\end{align}
	where  the amplitude of probe Rabi frequency $\Omega_0\ll\Gamma$ and $\tau$ is the x-ray pulse bandwidth.
	
	Since the incident x-ray pulse is very weak, we assume 	\begin{align}
	&\rho_{11}(z,t)-\rho_{44}(z,t)\approx 0.5\, , \label{equ12} \\
	&\rho_{22}(z,t)-\rho_{33}(z,t)\approx 0.5\, , \label{equ13} 
	\end{align}
	
	Inserting Eq.~\ref{equ12} and Eq.~\ref{equ13} into the MBEs and we obtain
	
	\begin{align}
	{{\partial }_{t}}{{\rho }_{32}}(z,t)=&-\frac{1}{2}(2i{{\Delta }_{p,3\to 2}}+C_{13}^{2}\Gamma +C_{23}^{2}\Gamma ){{\rho }_{32}}(z,t){+}\frac{i}{4}{{C}_{23}}{{\Omega }_{p}}(z,t) \, , \label{equ14} \\
	{{\partial }_{t}}{{\rho }_{41}}(z,t)=&-\frac{1}{2}(2i{{\Delta }_{p,4\to 1}}+C_{14}^{2}\Gamma +C_{24}^{2}\Gamma ){{\rho }_{41}}(z,t){+}\frac{i}{4}{{C}_{14}}{{\Omega }_{p}}(z,t) \, , \label{equ15} \\
	\frac{1}{c}\partial _t\Omega_{p}+\partial _z&\Omega_p=i\beta\left(\frac{\rho _{41}}{C_{14}}+\frac{\rho _{32}}{C_{23}}\right)\, . \label{equ16} 
	\end{align}
	Since $\Omega_0$ is real and $C_{13}^{2}+C_{23}^{2}=C_{14}^{2}+C_{24}^{2}=1,C_{23}=C_{14}=C$, the following equations are obtained depending on the symmetry
	\begin{align}
	\rho_{P}(z,t)=&{{\rho }_{41}}(z,t){+}{{\rho }_{32}}(z,t)=2i{Im}\left[{{\rho }_{41}}(z,t)\right]\, , \label{equ17} \\
	\rho_{S}(z,t)=&{{\rho }_{41}}(z,t)-{{\rho }_{32}}(z,t)=2{Re}\left[{{\rho }_{41}}(z,t)\right]\, . \label{equ18}  
	\end{align}
	Applying the definitions of $\rho_P$ and $\rho_S$ in Eq.~\ref{equ14} to ~\ref{equ15}, the following equations are derived
	\begin{align}
	{{\partial }_{t}}{{\rho }_{S}}(z,t)=&-\frac{\Gamma}{2}{{\rho }_{S}}(z,t)-i\Delta (t){{\rho }_{P}}(z,t) \, , \label{equ19} \\ 
	{{\partial }_{t}}{{\rho }_{P}}(z,t)=&-\frac{\Gamma}{2}{{\rho }_{P}}(z,t)-i\Delta (t){{\rho }_{S}}(z,t)+i\frac{C}{2}{{\Omega }_{p}(z,t)}\, ,\label{equ20}\\
	\frac{1}{c}\partial_t\Omega_{p}(z,t)&+{{\partial }_{z}}{{\Omega }_{p}(z,t)}=i\frac{\beta}{C}\rho_{P}(z,t) \, . 
	\label{equ21}
	\end{align}
	where 
	\begin{align}
	&\Delta (t)={{\Delta }_{p,4\to 1}}=-{{\Delta }_{p,3\to 2}}=\frac{1}{\hbar }\left({m}_{e4}{{\mu }_{e}}-{{m}_{g1}}{{\mu }_{g}}\right){{\mu }_{{N}}}{B}(t)\, .\label{equ22} 
	\end{align}
	Here, $\hbar$ is reduced Planck constant, ${m}_{e4}=-\frac{1}{2}$,${m}_{g1}=-\frac{1}{2}$, ${\mu }_{g}$ and ${\mu}_{e}$ are the Land\'e factors of the ground states and excited states, respectively; ${{\mu }_{{N}}}$ is nuclear magneton; $B(t)$ is the magnetic flux density.
	
Do the approximation that 
	\begin{equation}
	\frac{1}{c}\partial_t\Omega_{p}(z,t)\approx 0\, .\label{equ23} 
	\end{equation}
which is valid in our system and define 
\begin{align}
{{\rho }_{S}}(z,t)&=2{{\rho }_{R}}(z,t){{e}^{-\frac{\Gamma }{2}t}}\, ,
\,\,\,\,\,\,{{\rho }_{P}}(z,t)=2i{{\rho }_{I}}(z,t){{e}^{-\frac{\Gamma }{2}t}}\, .\label{equ24}
\end{align}

Finally we derive the simplified MBEs under weak and slowing varying approximations
	\begin{align}
	&{{\partial }_{t}}{{\rho }_{R}}(z,t)={{\rho }_{I}}(z,t)\Delta (t) \, , \label{equ25} \\ 
	&{{\partial }_{t}}{{\rho }_{I}}(z,t)=-{{\rho }_{R}}(z,t)\Delta (t)+\frac{C}{4}{{\Omega }_{p}}(z,t){{e}^{\frac{\Gamma }{2}t}} \, , \label{equ26} \\ 
	&{{\partial }_{z}}{{\Omega }_{p}}=-\frac{2}{C}\beta {{\rho }_{I}}(z,t){{e}^{-\frac{\Gamma }{2}t}} \, . \label{equ27} 
	\end{align}
	
	\section{2. The approximate analytical solution under the first magnetic pulse}
	In our protocol, we first consider the magnetic field has the same temporal shape as the input x-ray photon with the condition that $\text{A}_{B}=\int\Delta_{1}(t)dt\approx\pi$ and the first pulse magnetic arrives simultaneously as the single x-ray photon. Define 
	\begin{align}
	&\Delta_{1}(t)=\Delta_{01}{{e}^{-2\ln 2{{(\frac{t-{{t}_{0}}}{\tau })}^{2}}}}\, . \label{equ28} 
	\end{align}
	where $\Delta_{01}$ is  the splitting amplitude caused by the first magnetic field. In the following we present the analytical solutions of Eq.~\ref{equ25} to Eq.~\ref{equ27}.
	\begin{align}
	{{\Omega }_{p}}\left(z,t\right)&={{\Omega }_{p}}\left(0,t\right)-\frac{2}{C}\beta \int_{0}^{z}{{{\rho }_{I}}\left(z',t\right){{e}^{-\frac{\Gamma }{2}t}}dz'}\, , \label{equ29} \\
	{{\partial }_{t}}{{\rho }_{I}}\left(z,t\right)&=-{{\rho }_{R}}\left(z,t\right)\Delta \left(t\right)+\frac{C\Omega_{0}}{4\Delta_{01}}{{e}^{\frac{\Gamma }{2}t}}\Delta \left(t\right)-\frac{\beta }{2}\int_{0}^{z}{{{\rho }_{I}}\left(z',t\right)dz'}\, , \\
	&\approx -{{\rho }_{R}}\left(z,t\right)\Delta \left(t\right)+\frac{C\Omega_{0}}{4\Delta_{01}}{{e}^{\frac{\Gamma }{2}t_{0}}}\Delta \left(t\right)-\frac{\beta }{2}\int_{0}^{z}{{{\rho }_{I}}\left(z',t\right)dz'}\, .\label{equ30}
	\end{align}
	At z = 0, we have
	\begin{align}
	& {{\partial }_{t}}{{\rho }_{I}}\left(z,t\right)=-{{\rho }_{R}}\left(z,t\right)\Delta \left(t\right)+\frac{C\Omega_{0}}{4\Delta_{01}}{{e}^{\frac{\Gamma }{2}t_{0}}}\Delta \left(t\right)\, ,\label{32} \\ 
	& {{\partial }_{t}}{{\rho }_{R}}\left(z,t\right)={{\rho }_{I}}\left(z,t\right)\Delta \left(t\right)\, .\label{33} 
	\end{align}
	The analytical solution is derived 
	\begin{align}
	{{\rho }_{R}}\left(0,t\right)&={\frac{C\Omega_{0}}{4\Delta_{01}}{{e}^{\frac{\Gamma }{2}t_{0}}}}\left(1-\text{Cos} \left[\int_{t_{01}}^{t}{\Delta_{1} \left(t'\right)dt'}\right]\right) \, ,\label{34} \\ 
	{{\rho }_{I}}\left(0,t\right)&={\frac{C\Omega_{0}}{4\Delta_{01}}{{e}^{\frac{\Gamma }{2}t_{0}}}}\text{Sin} \left[\int_{t_{01}}^{t}{\Delta_{1} \left(t'\right)dt'}\right] \, .\label{35} 
	\end{align}
For $z\ne0$, we set ${{z}_{i}}=idz$ where $dz$ is infinitesimal.
	At $z={z}_{1}$, we have
	\begin{align}
	& {{\partial }_{t}}{{\rho }_{I}}\left(z,t\right)=-{{\rho }_{R}}\left(z,t\right)\Delta \left(t\right)+{\frac{C\Omega_{0}}{4\Delta_{01}}{{e}^{\frac{\Gamma }{2}t_{0}}}}\Delta \left(t\right)-\frac{\beta }{2}{\frac{C\Omega_{0}}{4\Delta_{01}}{{e}^{\frac{\Gamma }{2}t_{0}}}}\text{Sin} \left[\int_{t_{01}}^{t}{\Delta_{1} \left(t'\right)dt'}\right]dz \, , \label{equ36}\\ 
	& {{\partial }_{t}}{{\rho }_{R}}\left(z,t\right)={{\rho }_{I}}\left(z,t\right)\Delta \left(t\right) \, . \label{equ37}
	\end{align}	
	As $A_{B}=\pi$, an approximation is used as follows
	\begin{align}
	\text{Sin} \left[\int_{t_{01}}^{t}{\Delta_{1} \left(t'\right)dt'}\right]\approx\frac{\Delta_{1}(t)}{\Delta_{01}}\, .\label{equ38}
	\end{align}
Substituting Eq.~\ref{equ38} into Eq.~\ref{equ36} and Eq.~\ref{equ37}, we obtain
	\begin{align}
	& {{\partial }_{t}}{{\rho }_{I}}\left(z,t\right)=-{{\rho }_{R}}\left(z,t\right)\Delta \left(t\right)+{\frac{C\Omega_{0}}{4\Delta_{01}}{{e}^{\frac{\Gamma }{2}t_{0}}}}\left(1-\frac{\beta}{2\Delta_{01}}dz\right)\Delta \left(t\right) \, ,\label{39} \\
	& {{\partial }_{t}}{{\rho }_{R}}\left(z,t\right)={{\rho }_{I}}\left(z,t\right)\Delta \left(t\right) \, .\label{40} 
	\end{align}
Then we derive 
	\begin{align}
	& {{\rho }_{R}}\left({{z}_{1}},t\right)={\frac{C\Omega_{0}}{4\Delta_{01}}{{e}^{\frac{\Gamma }{2}t_{0}}}}\left(1-\frac{\beta}{2\Delta_{01}}dz\right)\left(1-\text{Cos} \left[\int_{t_{01}}^{t}{\Delta_{1} \left(t'\right)dt'}\right]\right) \, ,\label{41} \\ 
	& {{\rho }_{I}}\left({{z}_{1}},t\right)={\frac{C\Omega_{0}}{4\Delta_{01}}{{e}^{\frac{\Gamma }{2}t_{0}}}}\left(1-\frac{\beta}{2\Delta_{01}}dz\right)\text{Sin} \left[\int_{t_{01}}^{t}{\Delta_{1} \left(t'\right)dt'}\right] \, .\label{42} 
	\end{align}
For $z={{z}_{2}}$ and $z={{z}_{3}}$, it is easily derived that
	\begin{align}
	& {{\rho }_{R}}\left({{z}_{2}},t\right)={\frac{C\Omega_{0}}{4\Delta_{01}}{{e}^{\frac{\Gamma }{2}t_{0}}}}{{\left(1-\frac{\beta}{2\Delta_{01}}dz\right)}^{2}}\left(1-\text{Cos} \left[\int_{t_{01}}^{t}{\Delta_{1} \left(t'\right)dt'}\right]\right) \, ,\\ 
	& {{\rho }_{I}}\left({{z}_{2}},t\right)={\frac{C\Omega_{0}}{4\Delta_{01}}{{e}^{\frac{\Gamma }{2}t_{0}}}}{{\left(1-\frac{\beta}{2\Delta_{01}}dz\right)}^{2}}\text{Sin} \left[\int_{t_{01}}^{t}{\Delta_{1} \left(t'\right)dt'}\right] \, ,\\ 
	& {{\rho }_{R}}\left({{z}_{3}},t\right)={\frac{C\Omega_{0}}{4\Delta_{01}}{{e}^{\frac{\Gamma }{2}t_{0}}}}{{\left(1-\frac{\beta}{2\Delta_{01}}dz\right)}^{3}}\left(1-\text{Cos} \left[\int_{t_{01}}^{t}{\Delta_{1} \left(t'\right)dt'}\right]\right) \, ,\\ 
	& {{\rho }_{I}}\left({{z}_{3}},t\right)={\frac{C\Omega_{0}}{4\Delta_{01}}{{e}^{\frac{\Gamma }{2}t_{0}}}}{{\left(1-\frac{\beta}{2\Delta_{01}}dz\right)}^{3}}\text{Sin} \left[\int_{t_{01}}^{t}{\Delta_{1} \left(t'\right)dt'}\right] \, .
	\end{align}
For $z=z_n$, a recursion formula is presented in the following
	\begin{align}
	& {{\rho }_{R}}\left({{z}_{n}},t\right)={\frac{C\Omega_{0}}{4\Delta_{01}}{{e}^{\frac{\Gamma }{2}t_{0}}}}{{\left(1-\frac{\beta}{2\Delta_{01}}dz\right)}^{n}}\left(1-\text{Cos} \left[\int_{t_{01}}^{t}{\Delta_{1} \left(t'\right)dt'}\right]\right) \, ,\\ 
	& {{\rho }_{I}}\left({{z}_{n}},t\right)={\frac{C\Omega_{0}}{4\Delta_{01}}{{e}^{\frac{\Gamma }{2}t_{0}}}}{{\left(1-\frac{\beta}{2\Delta_{01}}dz\right)}^{n}}\text{Sin} \left[\int_{t_{01}}^{t}{\Delta_{1} \left(t'\right)dt'}\right] \, . 
	\end{align}
Using
	\begin{align}
	\underset{n\to +\infty }{\mathop{\lim }}\,{{\left(1-\frac{\beta}{2\Delta_{01}}dz\right)}^{n}}=\underset{n\to +\infty }{\mathop{\lim }}\,{{\left(1-\frac{\beta}{2\Delta_{01}}\frac{z}{n}\right)}^{n}}={{e}^{-\frac{\beta z}{2\Delta_{01}}}}\, ,
	\end{align}
we derive 
	\begin{align}
	& {{\rho }_{R}}\left(z,t\right)={\frac{C\Omega_{0}}{4\Delta_{01}}{{e}^{\frac{\Gamma }{2}t_{0}}}}{{e}^{-\frac{\beta z}{2\Delta_{01}}}}\left(1-\text{Cos} \left[\int_{t_{01}}^{t}{\Delta_{1} \left(t'\right)dt'}\right]\right)\, , \\ 
	& {{\rho }_{I}}\left(z,t\right)={\frac{C\Omega_{0}}{4\Delta_{01}}{{e}^{\frac{\Gamma }{2}t_{0}}}}{{e}^{-\frac{\beta z}{2\Delta_{01}}}}\text{Sin} \left[\int_{t_{01}}^{t}{\Delta_{1} \left(t'\right)dt'}\right]\, , 
	\end{align}
and 
	\begin{align}
	{{\Omega }^1_{p}}(z,t){=}{{\Omega }_{p}}(0,t)-\Omega_{0}\left(1-{e}^{-\frac{\beta z}{2\Delta_{01}}}\right)\text{Sin}\left[\int_{{{t}_{01}}}^{t}{{{\Delta }_{1}}(t')dt'}\right]{{e}^{-\frac{\Gamma }{2}(t-t_{0})}}\, .
	\end{align}
	
	\section{3. The approximate analytical solution under the second magnetic pulse}
	
Here, we consider the second magnetic field pulse which has the same temporal shape as the input x-ray photon. Before the arrival of the second magnetic pulse, we have
\begin{equation}
{{\rho }_{R}}\left(z,t\right)=2{\frac{C\Omega_{0}}{4\Delta_{01}}{{e}^{\frac{\Gamma }{2}t_{0}}}}{{e}^{-\frac{\beta z}{2\Delta_{01}}}}\,\,\, , {{\rho }_{I}}\left(z,t\right)=0\,\,\, , {{\Omega }_{p}}\left(z,t\right)=0 \, .
\end{equation}
Then the corresponding equations are
	\begin{align}
	& {{\partial }_{t}}{{\rho }_{I}}\left(z,t\right)=-{{\rho }_{R}}\left(z,t\right){{\Delta }_{2}}\left(t\right)-\frac{\beta }{2}\int_{0}^{z}{{{\rho }_{I}}\left(x,t\right)dx} \, ,\label{equ54}\\ 
	& {{\partial }_{t}}{{\rho }_{R}}\left(z,t\right)={{\rho }_{I}}\left(z,t\right){{\Delta }_{2}}\left(t\right) \, ,\label{equ55}\\ 
	& {{\partial }_{z}}{{\Omega }_{p}}\left(z,t\right){=}-\frac{2}{C}\beta {{\rho }_{I}}\left(z,t\right){{e}^{-\frac{\Gamma }{2}t}} \, .\label{equ56}
	\end{align}
At z=0,
	\begin{align}
	& {{\partial }_{t}}{{\rho }_{R}}\left(z,t\right)={{\rho }_{I}}\left(z,t\right){{\Delta }_{2}}\left(t\right)\, , \\
	& {{\partial }_{t}}{{\rho }_{I}}\left(z,t\right)=-{{\rho }_{R}}\left(z,t\right){{\Delta }_{2}}\left(t\right)\, .
	\end{align}
We derive 
	\begin{align}
	& {{\rho }_{R}}\left(0,t\right)={{\rho }_{{{R}_{1}}}}\left(0\right)\text{Cos} \left[\int_{{{t}_{02}}}^{t}{{{\Delta }_{2}}\left(t'\right)dt'}\right]\, , \\
	& {{\rho }_{I}}\left(0,t\right)=-{{\rho }_{{{R}_{1}}}}\left(0\right)\text{Sin} \left[\int_{{{t}_{02}}}^{t}{{{\Delta }_{2}}\left(t'\right)dt'}\right]\, . 
	\end{align}
The solution for $z\ne0$ is derived using the same method as described before. For $z=z_{1}$,
	\begin{align}
	{{\partial }_{t}}{{\rho }_{I}}\left(z,t\right)=&-{{\rho }_{R}}\left(z,t\right)\Delta_{2} \left(t\right)+\frac{\beta }{2}{{\rho }_{{{R}_{1}}}}\left(0\right)\text{Sin} \left(\int_{{{t}_{02}}}^{t}{\Delta_{2} \left(t'\right)dt'}\right)dz \nonumber \\ 
	\approx&-{{\rho }_{R}}\left(z,t\right)\Delta_{2} \left(t\right)+\frac{\beta }{2\Delta_{01}}{{\rho }_{{{R}_{1}}}}\left(0\right)\Delta_{2}(t)dz\, , \\ 
	{{\partial }_{t}}{{\rho }_{R}}\left(z,t\right)=&{{\rho }_{I}}\left(z,t\right)\Delta_{2} \left(t\right)\, . 
	\end{align}
The derived solution is presented as follows
	\begin{align}
	& {{\rho }_{R}}\left({{z}_{1}},t\right)=\left({{\rho }_{{{R}_{1}}}}\left({{z}_{1}}\right)-{{\rho }_{{{R}_{1}}}}\left(0\right)\frac{\beta }{2\Delta_{01}}dz\right)\text{Cos} \left[\int_{{{t}_{02}}}^{t}{\Delta_{2} \left(t'\right)dt'}\right]+{{\rho }_{{{R}_{1}}}}\left(0\right)\frac{\beta }{2\Delta_{01}}dz\, , \\ 
	& {{\rho }_{I}}\left({{z}_{1}},t\right)=-\left({{\rho }_{{{R}_{1}}}}\left({{z}_{1}}\right)-{{\rho }_{{{R}_{1}}}}\left(0\right)\frac{\beta }{2\Delta_{01}}dz\right)\text{Sin} \left[\int_{{{t}_{02}}}^{t}{\Delta_{2} \left(t'\right)dt'}\right]\, .
	\end{align}
We define
	\begin{align}
	& {{F}^{\left(0\right)}}\left(z\right)={{\rho }_{{{R}_{1}}}}\left(z\right)\, , \\ 
	& {{F}^{\left(1\right)}}\left(z\right)=\int_{0}^{z}{{{\rho }_{{{R}_{1}}}}\left({{z}^1}\right)}d{{z}^1}\, , \\ 
	& {{F}^{\left(2\right)}}\left(z\right)=\int_{0}^{z}{\int_{0}^{{{z}^1}}{{{\rho }_{{{R}_{1}}}}\left({{z}^2}\right)}}d{{z}^2}d{{z}^1}\, ,\\ 
	& {{F}^{\left(n\right)}}\left(z\right)=\int_{0}^{z}{\cdots \int_{0}^{{{z}^{n-1}}}{{{\rho }_{{{R}_{1}}}}\left({{z}^{n}}\right)}}d{{z}^{n}}\cdots d{{z}^{1}}\, . 
	\end{align}
The solution is derived 
	\begin{align}
	& {{\rho }_{R}}\left(z,t\right)=\underset{n\to +\infty }{\mathop{\lim }}\,\sum\limits_{j=0}^{n}{{{\left(-\frac{\beta z}{2\Delta_{01}}\right)}^{j}}}{{F}^{\left(j\right)}}\left(z\right)\text{Cos} \left[\int_{{{t}_{02}}}^{t}{\Delta_{2} \left(t'\right)dt'}\right]-\underset{
		n\to +\infty}{\mathop{\lim}}\,\sum\limits_{j=1}^{n}{{{\left(-\frac{\beta z}{2\Delta_{01}}\right)}^{j}}}{{F}^{\left(j\right)}}\left(z\right)\, ,\\ 
	& {{\rho }_{I}}\left(z,t\right)=-\underset{n\to +\infty }{\mathop{\lim }}\,\sum\limits_{j=0}^{n}{{{\left(-\frac{\beta z}{2\Delta_{01}}\right)}^{j}}}{{F}^{\left(j\right)}}\left(z\right)\text{Sin} \left[\int_{{{t}_{02}}}^{t}{\Delta_{2} \left(t'\right)dt'}\right] \, .\\ 
	\end{align}
In the following we go into details of ${{\left(-\frac{\beta z}{2\Delta_{01}}\right)}^{n}}{{F}^{\left(n\right)}}\left(z\right)$:
	\begin{align}
	{{F}^{\left(0\right)}}\left(z\right)&=\left({\frac{C\Omega_{0}}{2\Delta_{01}}{{e}^{\frac{\Gamma }{2}t_{0}}}}\right){{e}^{-\frac{\beta z}{2\Delta_{01}}}}\, , \\ 
	\left(-\frac{\beta z}{2\Delta_{01}}\right){{F}^{\left(1\right)}}\left(z\right)&=\left({\frac{C\Omega_{0}}{2\Delta_{01}}{{e}^{\frac{\Gamma }{2}t_{0}}}}\right)\left({{e}^{-\frac{\beta z}{2\Delta_{01}}}}-1\right)\, , \\ 
	{{\left(-\frac{\beta z}{2\Delta_{01}}\right)}^{2}}{{F}^{\left(2\right)}}\left(z\right)&=\left({\frac{C\Omega_{0}}{2\Delta_{01}}{{e}^{\frac{\Gamma }{2}t_{0}}}}\right)\left[{{e}^{-\frac{\beta z}{2\Delta_{01}}}}-\left(1-\frac{\beta z}{2\Delta_{01}}\right)\right]\, , \\ 
	{{\left(-\frac{\beta z}{2\Delta_{01}}\right)}^{n}}{{F}^{\left(n\right)}}\left(z\right)&=\left({\frac{C\Omega_{0}}{2\Delta_{01}}{{e}^{\frac{\Gamma }{2}t_{0}}}}\right)\left[{{e}^{-\frac{\beta z}{2\Delta_{01}}}}-\sum\limits_{j=0}^{n-1}{\frac{1}{j!}{{\left(-\frac{\beta z}{2\Delta_{01}}\right)}^{j}}}\right]\, .
	\end{align}
Then we have 
	\begin{align}
	& \sum\limits_{j=0}^{n}{{{\left(-\frac{\beta z}{2\Delta_{01}}\right)}^{j}}}{{F}^{\left(j\right)}}\left(z\right)/\left({\frac{C\Omega_{0}}{2\Delta_{01}}{{e}^{\frac{\Gamma }{2}t_{0}}}}\right){=}\left(n+1\right){{e}^{-\frac{\beta z}{2\Delta_{01}}}}-\sum\limits_{j=0}^{n-1}{\frac{n-j}{\left(j\right)!}{{\left(-\frac{\beta z}{2\Delta_{01}}\right)}^{j}}}\, ,\\
	&={{e}^{-\frac{\beta z}{2\Delta_{01}}}}+\left(-\frac{\beta z}{2\Delta_{01}}\right)\sum\limits_{j=0}^{n-2}{\frac{1}{\left(n-2\right)!}{{\left(-\frac{\beta z}{2\Delta_{01}}\right)}^{n-2}}}+n\sum\limits_{j=n}^{+\infty}{\frac{1}{j!}{{\left(-\frac{\beta z}{2\Delta_{01}}\right)}^{j}}}\, .
	\end{align}
When $n\to +\infty $, we obatin
	\begin{align}
	\underset{n\to +\infty }{\mathop{\lim }}\,\sum\limits_{j=0}^{n}{{{\left(-\frac{\beta z}{2\Delta_{01}}\right)}^{j}}}{{F}^{\left(j\right)}}\left(z\right){=}\left({\frac{C\Omega_{0}}{2\Delta_{01}}{{e}^{\frac{\Gamma }{2}t_{0}}}}\right)\left(1-\frac{\beta z}{2\Delta_{01}}\right){{e}^{-\frac{\beta z}{2\Delta_{01}}}}\, .
	\end{align}
Then we derive the final expressions
	\begin{align}
	{{\rho }_{R}}\left(z,t\right)&=\left({\frac{C\Omega_{0}}{2\Delta_{01}}{{e}^{\frac{\Gamma }{2}t_{0}}}}\right)\left(1-\frac{\beta z}{2\Delta_{01}}\right){{e}^{-\frac{2\Gamma\xi z}{\Delta_{01}L}}}\text{Cos} \left[\int_{{{t}_{02}}}^{t}{{{\Delta }_{2}}\left(t'\right)dt'}\right]+\left({\frac{C\Omega_{0}}{2\Delta_{01}}{{e}^{\frac{\Gamma }{2}t_{0}}}}\right)\frac{\beta z}{2\Delta_{01}}{{e}^{-\frac{\beta z}{2\Delta_{01}}}}\, ,\\ 
	{{\rho }_{I}}\left(z,t\right)&=-\left({\frac{C\Omega_{0}}{2\Delta_{01}}{{e}^{\frac{\Gamma }{2}t_{0}}}}\right)\left(1-\frac{\beta z}{2\Delta_{01}}\right){{e}^{-\frac{\beta z}{2\Delta_{01}}}}\text{Sin} \left[\int_{{{t}_{02}}}^{t}{{{\Delta }_{2}}\left(t'\right)dt'}\right]\, ,\\ 
	{{\Omega }^2_{p}}(z,t)&=-\frac{2}{C}\beta e^{-\frac{\Gamma}{2}t}\int_{0}^{z}\rho_{I}(z',t)dz'\nonumber\\
	&=2{{\Omega }_{0}}\frac{\beta z}{2\Delta_{01}}{{e}^{-\frac{\beta z}{2\Delta_{01}}}}\text{Sin}\left[\int_{{{t}_{02}}}^{t}{{{\Delta }_{2}}(t')dt'}\right]{{e}^{-\frac{\Gamma }{2}(t-t_{0})}}\, . 
	\end{align}
	
	\section{4. The approximate analytical solution under the third magnetic pulse}
	Next,we consider the third magnetic field pulse which has the same temporal shape as the input x-ray photon.Before the arrival of the third magnetic pulse,we have
	\begin{align}
	{{\rho }_{R}}\left(z,t\right)={{\rho }_{{{R}_{2}}}}\left(z\right)=\left({\frac{C\Omega_{0}}{2\Delta_{01}}{{e}^{\frac{\Gamma }{2}t_{0}}}}\right)\left(2\frac{\beta z}{2\Delta_{01}}-1\right){{e}^{-\frac{\beta z}{2\Delta_{01}}}}\,\,\, ,{{\rho }_{I}}\left(z,t\right)=0\,\,\, ,{{\Omega }_{p}}\left(z,t\right)=0 \, .
	\end{align}
	And just like we did for the second magnetic pulse, we can define
	\begin{align}
	& {{F}_{3}^{\left(0\right)}}\left(z\right)={{\rho }_{{{R}_{2}}}}\left(z\right)\, ,\\ 
	& {{F}_{3}^{\left(1\right)}}\left(z\right)=\int_{0}^{z}{{{\rho }_{{{R}_{2}}}}\left({{x}_{1}}\right)}d{{x}_{1}}\, , \\ 
	& {{F}_{3}^{\left(2\right)}}\left(z\right)=\int_{0}^{z}{\int_{0}^{{{z}^{1}}}{{{\rho }_{{{R}_{2}}}}\left({{z}^{2}}\right)}}d{{z}^{2}}d{{z}^{1}}\, , \\ 
	& {{F}_{3}^{\left(n\right)}}\left(z\right)=\int_{0}^{z}{\cdots \int_{0}^{{{z}^{n-1}}}{{{\rho }_{{{R}_{2}}}}\left({{z}^{n}}\right)}}d{{z}^{n}}\cdots d{{z}^{1}}\, . 
	\end{align}
	the solution at z can be
	\begin{align}
	& {{\rho }_{R}}\left(z,t\right)=\underset{n\to +\infty }{\mathop{\lim }}\,\sum\limits_{j=0}^{n}{{{\left(-\frac{\beta z}{2\Delta_{01}}\right)}^{j}}}{{F}_{3}^{\left(j\right)}}\left(z\right)\text{Cos} \left[\int_{{{t}_{03}}}^{t}{\Delta_{3} \left(t'\right)dt'}\right]\, ,\\ 
	& {{\rho }_{I}}\left(z,t\right)=-\underset{n\to +\infty }{\mathop{\lim }}\,\sum\limits_{j=0}^{n}{{{\left(-\frac{\beta z}{2\Delta_{01}}\right)}^{j}}}{{F}_{3}^{\left(j\right)}}\left(z\right)\text{Sin} \left[\int_{{{t}_{03}}}^{t}{\Delta_{3} \left(t'\right)dt'}\right]\, . 
	\end{align}
	it is easy to get the general formula of ${{\left(-\frac{\beta z}{2\Delta_{01}}\right)}^{n}}{{F}^{\left(n\right)}}\left(z\right)$
	\begin{align}
	{{\left(-\frac{\beta z}{2\Delta_{01}}\right)}^{n}}{{F}_{3}^{\left(n\right)}}\left(z\right)/\left({\frac{C\Omega_{0}}{2\Delta_{01}}{{e}^{\frac{\Gamma }{2}t_{0}}}}\right)&=-\left(2n-1+2\frac{\beta z}{2\Delta_{01}}\right){{e}^{-\frac{\beta z}{2\Delta_{01}}}}+\sum\limits_{j=0}^{n-1}{\frac{2n-1-2j}{j!}{{\left(-\frac{\beta z}{2\Delta_{01}}\right)}^{j}}}\, .
	\end{align}
	Then we derive
	\begin{align}
	& \sum\limits_{j=0}^{n}{{{\left(-\frac{\beta z}{2\Delta_{01}}\right)}^{j}}}{{F}_{3}^{\left(j\right)}}\left(z\right)/\left({\frac{C\Omega_{0}}{2\Delta_{01}}{{e}^{\frac{\Gamma }{2}t_{0}}}}\right){=}\left(1-n^{2}-2\left(n-1\right)\frac{\beta z}{2\Delta_{01}}\right){{e}^{-\frac{\beta z}{2\Delta_{01}}}}-\sum\limits_{j=0}^{n-1}{\frac{(n-j)^{2}}{\left(j\right)!}{{\left(-\frac{\beta z}{2\Delta_{01}}\right)}^{j}}}\, .
	\end{align}
	When $n\to +\infty $,we obtain
	\begin{align}
	\underset{n\to +\infty }{\mathop{\lim }}\,\sum\limits_{j=0}^{n}{{{\left(-\frac{\beta z}{2\Delta_{01}}\right)}^{j}}}{{F}_{3}^{\left(j\right)}}\left(z\right){=}\left({\frac{C\Omega_{0}}{2\Delta_{01}}{{e}^{\frac{\Gamma }{2}t_{0}}}}\right)\left(1-3\frac{\beta z}{2\Delta_{01}}+\left(\frac{\beta z}{2\Delta_{01}}\right)^{2}\right){{e}^{-\frac{\beta z}{2\Delta_{01}}}}\, .
	\end{align}
	The corresponding solution to the third magnetic pulse
	\begin{align}
	{{\rho }_{R}}\left(z,t\right)&=-\left({\frac{C\Omega_{0}}{2\Delta_{01}}{{e}^{\frac{\Gamma }{2}t_{0}}}}\right)\left(1-3\frac{\beta z}{2\Delta_{01}}+\left(\frac{\beta z}{2\Delta_{01}}\right)^{2}\right){{e}^{-\frac{\beta z}{2\Delta_{01}}}}\text{Cos} \left[\int_{{{t}_{03}}}^{t}{{{\Delta }_{3}}\left(t'\right)dt'}\right]\nonumber\\
	&+\left(-\frac{2\Gamma\xi z}{\Delta_{01}L}+\left(\frac{\beta z}{2\Delta_{01}}\right)^{2}\right){{e}^{-\frac{\beta z}{2\Delta_{01}}}}\, ,\\ 
	{{\rho }_{I}}\left(z,t\right)&=\left({\frac{C\Omega_{0}}{2\Delta_{01}}{{e}^{\frac{\Gamma }{2}t_{0}}}}\right)\left(1-3\frac{\beta z}{2\Delta_{01}}+\left(\frac{\beta z}{2\Delta_{01}}\right)^{2}\right){{e}^{-\frac{\beta z}{2\Delta_{01}}}}\text{Sin} \left[\int_{{{t}_{03}}}^{t}{{{\Delta }_{3}}\left(t'\right)dt'}\right]\, ,\\ 
	{{\Omega }^3_{p}}(z,t)&=-\frac{2}{C}\beta e^{-\frac{\Gamma}{2}t}\int_{0}^{z}\rho_{I}(z',t)dz'\nonumber\\
	&=2{{\Omega }_{0}}\left(-\frac{\beta z}{2\Delta_{01}}+\left(\frac{\beta z}{2\Delta_{01}}\right)^{2}\right){{e}^{-\frac{\beta z}{2\Delta_{01}}}}\text{Sin}\left[\int_{{{t}_{03}}}^{t}{{{\Delta }_{3}}(t')dt'}\right]{{e}^{-\frac{\Gamma }{2}(t-t_{0})}}\, .
	\end{align}
	
	\section{5. General solution under a train of the magnetic field pulses}
	Using the recursive method, we assume the general solution of $\rho_R$ and $\rho_I$ under the $n$th magnetic field pulse is
	\begin{align}
	&\rho_{R}(z,t)=(-1)^{n}\left({\frac{C\Omega_{0}}{2\Delta_{01}}{{e}^{\frac{\Gamma }{2}t_{0}}}}\right)g(z)\text{Cos} \left[\int_{{{t}_{0n}}}^{t}{{{\Delta }_{n}}\left(t'\right)dt'}\right]+(-1)^{n-1}\left({\frac{C\Omega_{0}}{2\Delta_{01}}{{e}^{\frac{\Gamma }{2}t_{0}}}}\right)f(z) \, ,\label{equ98}\\
	&\rho_{I}(z,t)=(-1)^{n-1}\left({\frac{C\Omega_{0}}{2\Delta_{01}}{{e}^{\frac{\Gamma }{2}t_{0}}}}\right)g(z)\text{Sin} \left[\int_{{{t}_{0n}}}^{t}{{{\Delta }_{n}}\left(t'\right)dt'}\right] \, .\label{equ99}
	\end{align}
	Substituting Eq.~\ref{equ98}-Eq.~\ref{equ99} into Eq.~\ref{equ54}-Eq.~\ref{equ55}, we have
	\begin{align}
	\frac{dg(z)}{dz}=-\frac{\beta}{2\Delta_{01}}\left[g(z)+f(z)\right] \, .\label{equ100}
	\end{align}
	Using Eq.~\ref{equ100} and Eq.~\ref{equ54}-Eq.~\ref{equ55}, we obtain
	\begin{align}
	\Omega^{n}_{p}(z,t)=2\Omega_{0}(-1)^{n-1}e^{-\frac{\beta z}{2\Delta_{01}}}\sum\limits_{j=1}^{n-1}\frac{2^{j}}{j!}\left(-\frac{\beta z}{2\Delta_{01}}\right)^{j}F(n,j)\text{Sin} \left[\int_{{{t}_{0n}}}^{t}{{{\Delta }_{n}}\left(t'\right)dt'}\right]{{e}^{-\frac{\Gamma }{2}(t-t_{0})}} \, .\label{equ101}
	\end{align}
	where $F(n,j)$ is defined as 
	\begin{equation}
	F(n,j)=\left\{
	\begin{aligned}
	&1,j=1 \, , \\
	&n-j,j=2 \, , \\
	&\sum_{x_{j-2}=1}^{n-j}\ \sum_{x_{j-3}=1}^{x_{j-2}}\cdots\sum_{x_{1}=1}^{x_{2}}x_{1},j>2 \, .
	\end{aligned}
	\right.
	\end{equation}
	We have derived the general solution under the $n$th magnetic field pulse.

\end{widetext}

\end{document}